\documentclass[11pt]{article}
\usepackage[latin9]{inputenc}
\usepackage{amsfonts}
\usepackage{amsmath}
\usepackage{amssymb}
\usepackage{indentfirst}
\usepackage{graphicx}
\usepackage[colorlinks]{hyperref}
\usepackage{cite}

\numberwithin{equation}{section}
\begin{document}

\begin{titlepage}
\vspace{3cm}
\baselineskip=24pt

\begin{center}
\textbf{\LARGE{}
$\mathcal{N}$-extended Maxwell supergravities as Chern-Simons theories in three spacetime dimensions}%

\par\end{center}{\LARGE \par}

\begin{center}
	\vspace{1cm}
	\textbf{%
Patrick Concha}$^{\ddag}$
	\small
	\\[5mm]
	$^{\ddag}$\textit{Instituto
		de Física, Pontificia
Universidad Católica de Valparaíso, }\\
	\textit{ Casilla 4059,
Valparaiso-Chile.}
	\\[5mm]
	\footnotesize
	\texttt{patrick.concha@pucv.cl}
	\par\end{center}
\vskip 30pt
\begin{abstract}
\noindent
We present a new class of three-dimensional $\mathcal{N}$-extended supergravity theories based on the $\mathcal{N}$-extended Maxwell superalgebra with central charges and $\mathfrak{so}(\mathcal{N})$ internal symmetry generators. The presence of $\mathfrak{so}(\mathcal{N})$ generators is required in order to define a non-degenerate invariant inner product. Such symmetry allows us to construct an alternative supergravity action without cosmological constant term. Interestingly, the new theories can be obtained as a flat limit of a $\mathcal{N}$-extended AdS-Lorentz supergravity theories enlarged with $\mathfrak{so}(\mathcal{N})$ gauge fields.
\end{abstract}
\end{titlepage}\newpage {}

\section{Introduction}

It is well assumed that a three-dimensional (super)gravity theory can be
described by a Chern-Simons (CS) action as a gauge theory offering us an
interesting toy model to approach higher-dimensional theories \cite{DK,
Deser, Nieuwenhuizen, AT, RN, Witten, AT2, NG, HIPT, BTZ, GTW}. There has
been a growing interest to go beyond Poincaré and AdS (super)groups to
describe (super)gravity theories in order to study new models with different
physical content \cite{Sorokas, SS, DKGS, DFIMRSV, SalSal, GRSS, FMT, FMT2,
Durka, BDR, CCFRS, FISV, CDMR, CMRSV, Valcarcel}. In a previous work \cite%
{CPR}, we presented a new class of three-dimensional supergravity theories
based on the Maxwell and AdS-Lorentz superalgebras.

At the bosonic level, the Maxwell symmetry has been initially introduced in
\cite{BCR, Schrader, GK} in order to describe a Minkowski space in presence
of a constant classical electromagnetic field background. Its generalization
has been useful to recover General Relativity from CS and Born-Infeld (BI)
gravity actions \cite{CPRS1, CPRS2, CPRS3}. Recently, there has been a
particular interest in exploring CS gravity model invariant under the
Maxwell algebra \cite{SSV, HR, AFGHZ, CMMRSV}. In particular, the presence
of the additional gauge field influences the vacuum energy and the vacuum
angular momentum of the stationary configuration \cite{CMMRSV}.

On the other hand, the AdS-Lorentz symmetry was first introduced in \cite%
{SS, GKL} and can be seen as a semi-simple enlargement of the Poincaré
symmetry. Recently, the AdS-Lorentz algebra and its generalizations have
been used to recover diverse (pure)Lovelock Lagrangian from CS and BI ones
\cite{CDIMR, CMR, CR3}. More recently, it has been showed that the
asymptotic symmetry of a three-dimensional CS gravity action invariant under
the AdS-Lorentz group is given by a semi-simple enlargement of the $%
\mathfrak{bms}_{3}$ symmetry which is isomorphic to three copies of the
Virasoro algebra \cite{CMRSV}.

At the supersymmetric level, the Maxwell superalgebra has been introduced to
describe the presence of a constant abelian supersymmetric gauge field
background in a four-dimensional superspace \cite{BGKL}. Interestingly, a
pure supergravity action in four dimensions can be constructed based on the
Maxwell superalgebra using a geometric procedure \cite{AI, CR2}. Recently it
has been shown in \cite{CRR} that the bosonic extra field and the additional
Majorana gauge field, appearing in the Maxwell superalgebra, are crucial to
recover supersymmetry invariance of flat supergravity on a manifold with
non-trivial boundary. Extensions and generalizations of the supersymmetric
version of the Maxwell symmetries have been extensively studied by diverse
authors \cite{GKL2, BGKL2, Lukierski, FL, CK, CFRS, CFR, PR, Ravera, KSC, KC}%
.

In this paper we extend the results of \cite{CPR} and present a new class of
$\mathcal{N}$-extended CS supergravity theories in three spacetime
dimensions based on the central $\mathcal{N}$-extended Maxwell superalgebra.
We show that the Maxwell superalgebra has to be enlarged with $\mathfrak{so}%
\left( \mathcal{N}\right) $ generators in order to have non-degenerate
invariant inner product. Interestingly, the Maxwell superalgebra allows us
to define an alternative supergravity model in absence of cosmological
constant term. The introduction of the cosmological constant term is done by
considering an enlarged superalgebra which corresponds to a $\mathcal{N}$%
-extended AdS-Lorentz superalgebra. In order to establish a well-defined
Maxwell limit we extend the superalgebra with $\mathfrak{so}\left( \mathcal{N%
}\right) $ internal symmetry algebra.

This work is organized as follows: In Section 2, we give a brief review of
the three-dimensional minimal Maxwell CS supergravity theory. The section 3
is devoted to the construction of the CS supergravity action invariant under
the $\mathcal{N}=2$ Maxwell superalgebra. In section 4, we present a new
class of $\mathcal{N}$-extended supergravity models expressed as CS action
for the $\mathcal{N}$-extended Maxwell superalgebra. The obtention of the $%
\mathcal{N}$-extended Maxwell supergravity action through a Maxwell limit is
explored in Section 5. In Section 6, we give a brief discussion about future
possible developments.

\section{Minimal Maxwell Chern-Simons supergravity}

In this section, following \cite{CPR}, we give a brief review of the minimal
Maxwell supergravity theory based on the CS formalism. The minimal
supersymmetric extension of the Maxwell algebra in three spacetime
dimensions is spanned by the set $\left\{ J_{a},P_{a},Z_{a},Q_{a},\Sigma
_{\alpha }\right\} $ whose generators satisfy the following non-vanishing
(anti-)commutation relations:%
\begin{eqnarray}
\left[ J_{a},J_{b}\right] &=&\epsilon _{abc}J^{c}\,,\qquad \left[ J_{a},P_{b}%
\right] =\epsilon _{abc}P^{c}\,,  \notag \\
\left[ J_{a},Z_{b}\right] &=&\epsilon _{abc}Z^{c}\,,\qquad \left[ P_{a},P_{b}%
\right] =\epsilon _{abc}Z^{c}\,,  \notag \\
\left[ J_{a},Q_{\alpha }\right] &=&\frac{1}{2}\,\left( \Gamma _{a}\right) _{%
\text{ }\alpha }^{\beta }Q_{\beta }\,,\text{ \ \ }  \notag \\
\left[ J_{a},\Sigma _{\alpha }\right] &=&\frac{1}{2}\,\left( \Gamma
_{a}\right) _{\text{ }\alpha }^{\beta }\Sigma _{\beta }\,,\text{ \ \ }
\label{sp1} \\
\left[ P_{a},Q_{\alpha }\right] &=&\frac{1}{2}\,\left( \Gamma _{a}\right) _{%
\text{ }\alpha }^{\beta }\Sigma _{\beta }\,,\text{ }  \notag \\
\left\{ Q_{\alpha },Q_{\beta }\right\} &=&-\frac{1}{2}\left( C\Gamma
^{a}\right) _{\alpha \beta }P_{a}\,,  \notag \\
\left\{ Q_{\alpha },\Sigma _{\beta }\right\} &=&-\frac{1}{2}\,\left( C\Gamma
^{a}\right) _{\alpha \beta }Z_{a}\,,  \notag
\end{eqnarray}%
where the Lorentz indices $a,b,\cdots =0,1,2$ are lowered and raised with
the Minkowski metric $\eta _{ab}$. Here, $C$ is the charge conjugation
matrix,%
\begin{equation}
C_{\alpha \beta }=C^{\alpha \beta }=\left(
\begin{array}{cc}
0 & -1 \\
1 & 0%
\end{array}%
\right) \,\text{,}
\end{equation}%
and satisfies $C^{T}=-C$ and $C\Gamma ^{a}=\left( C\Gamma ^{a}\right) ^{T}$
where $\Gamma ^{a}$ are the Dirac matrices in three spacetime dimensions.

Such Maxwell superalgebra has been first introduced in \cite{BGKL} in four
spacetime dimensions and subsequently in three dimensions \cite{BGKL2}.
Although the supersymmetrization of the Maxwell symmetry is not unique \cite%
{Sorokas, Lukierski, CFRS, CFR}, this is the minimal supersymmetric
extension providing us a consistent three-dimensional CS supergravity action
\cite{CPR}. Their (anti-)commutators differ from those of the super Poincaré
ones in the presence of the abelian generators $Z_{a}$ and the additional
Majorana spinor generators $\Sigma _{\alpha }$. On the other hand, the
introduction of a second abelian spinors charges has been first considered
in the context of $D=11$ supergravity \cite{AF} and superstring theory \cite%
{Green}. Here, the second spinorial generator assure that the Jacobi
identities hold.

A CS action,%
\begin{equation}
I_{CS}=\frac{k}{4\pi }\,\int_{M}\left\langle AdA+\frac{2}{3}%
A^{3}\right\rangle \,,  \label{CSA}
\end{equation}%
invariant under the Maxwell superalgebra has been explicitly constructed in
\cite{CPR} using the gauge connection one-form $A=A_{\mu }dx^{\mu }$ and the
corresponding invariant tensor. In particular, the connection one-form is
given by%
\begin{equation}
A=\omega ^{a}J_{a}+e^{a}P_{a}+\sigma ^{a}Z_{a}+\bar{\psi}Q+\bar{\xi}\Sigma
\,,  \label{of}
\end{equation}%
where $\omega ^{a}$ corresponds to the spin connection, $e^{a}$ is the
vielbein, $\sigma ^{a}$ is the so-called gravitational Maxwell gauge field
\cite{CMMRSV} while $\psi $ and $\xi $ are the respective fermionic gauge
fields.

In order to construct the CS supergravity action we shall require the
following non-vanishing components of the invariant tensor \cite{CPR},%
\begin{eqnarray}
\left\langle J_{a}J_{b}\right\rangle &=&\alpha _{0}\eta _{ab}\,,\text{\qquad
}\left\langle P_{a}P_{b}\right\rangle =\alpha _{2}\eta _{ab}\,,  \notag \\
\left\langle J_{a}P_{b}\right\rangle &=&\alpha _{1}\eta _{ab}\,,\text{\qquad
}\left\langle Q_{\alpha }Q_{\beta }\right\rangle =\alpha _{1}C_{\alpha \beta
}\,,  \label{InvTensMax} \\
\left\langle J_{a}Z_{b}\right\rangle &=&\alpha _{2}\eta _{ab}\,,\text{\qquad
}\left\langle Q_{\alpha }\Sigma _{\beta }\right\rangle =\alpha _{2}C_{\alpha
\beta }\,,  \notag
\end{eqnarray}%
where $\alpha _{0}\,,\alpha _{1}$ and $\alpha _{2}$ are arbitrary constants.
Then, using the connection one-form (\ref{of}) and the non-vanishing
components of the invariant tensor (\ref{InvTensMax}), the explicit form of
the CS supergravity action reads,%
\begin{eqnarray}
I &=&\frac{k}{4\pi }\int \alpha _{0}\left( \,\omega ^{a}d\omega _{a}+\frac{1%
}{3}\,\epsilon _{abc}\omega ^{a}\omega ^{b}\omega ^{c}\right)  \notag \\
&&+\alpha _{1}\left( 2e^{a}R_{a}-\bar{\psi}\nabla \psi \right) \,+\alpha
_{2}\left( 2R^{a}\sigma _{a}+e^{a}T_{a}-\bar{\psi}\nabla \xi -\bar{\xi}%
\nabla \psi \right) \,,  \label{sMCS}
\end{eqnarray}%
where the Lorentz curvature $R^{a}$, the torsion $T^{a}$ and the fermionic
curvatures are respectively given by%
\begin{eqnarray}
R^{a} &=&d\omega ^{a}+\frac{1}{2}\epsilon ^{abc}\omega _{b}\omega _{c}\,,
\notag \\
T^{a} &=&de^{a}+\epsilon ^{abc}\omega _{b}e_{c}\,,  \notag \\
\nabla \psi &=&d\psi +\frac{1}{2}\,\omega ^{a}\Gamma _{a}\psi \,, \\
\nabla \xi &=&d\xi +\frac{1}{2}\,\omega ^{a}\Gamma _{a}\xi \,+\frac{1}{2}%
\,e^{a}\Gamma _{a}\psi \,.  \notag
\end{eqnarray}%
The Maxwell gravitational field $\sigma ^{a}$ and the additional Majorana
spinor field $\xi $ have only contribution on the exotic part ($\alpha _{2}$%
) of the CS action. This is due to the presence of the non-vanishing
component $\left\langle J_{a}Z_{b}\right\rangle $, $\left\langle
P_{a}P_{b}\right\rangle $ and $\left\langle Q_{\alpha }\Sigma _{\beta
}\right\rangle $ of the invariant tensor which do not appear for the Poincaré
superalgebra. Interestingly, as was noticed at the bosonic level in \cite%
{CMMRSV}, the vacuum energy and the vacuum angular momentum of the
stationary configuration are influenced by the presence of the gravitational
Maxwell field $\sigma ^{a}$. Here one can see that, for $\alpha _{2}\neq 0$,
the field equations reduce to the vanishing of the curvature two-forms,%
\begin{equation}
\begin{tabular}{lll}
$R^{a}=0\,,$ & $\mathcal{T}^{a}=0\,,$ & $\mathcal{F}^{a}=0\,,$ \\
$\nabla \psi =0\,,$ & $\nabla \xi =0\,,$ &
\end{tabular}%
\end{equation}%
where%
\begin{eqnarray}
\mathcal{T}^{a} &=&T^{a}+\frac{1}{4}\bar{\psi}\Gamma ^{a}\psi \,, \\
\mathcal{F}^{a} &=&d\sigma ^{a}+\epsilon ^{abc}\omega _{b}\sigma _{c}+\frac{1%
}{2}\epsilon ^{abc}e_{b}e_{c}+\frac{1}{2}\bar{\psi}\Gamma ^{a}\xi \,.
\end{eqnarray}

The $\mathcal{N}$-extension of generalized Maxwell superalgebras have been
already presented in \cite{AILW, CR1}. Such $\mathcal{N}$-extended
superalgebras contain not only internal symmetry generators but also require
the presence of additional bosonic generators different to the Maxwell one.
However, as we shall see in the next sections, the construction of a proper
CS supergravity action based on a $\mathcal{N}$-extended Maxwell
superalgebra will only require to introduce $\mathfrak{so}\left( \mathcal{N}%
\right) $ generators.

\section{Chern-Simons formulation of $\mathcal{N}=2$ Maxwell supergravity}

The $\mathcal{N}=2$ supersymmetric extension of the Maxwell algebra is not
unique and can be subdivides into two inequivalent cases: the $\left(
1,1\right) $ and the $\left( 2,0\right) $ cases. Here we shall refer to the $%
\mathcal{N}=2$ Maxwell theory as $\left( 2,0\right) $ Maxwell supergravity
theory.

The extension to $\mathcal{N}\geq 2$ of the Maxwell superalgebra allows us
to include a central charge $Z$ to the usual Maxwell generators $\left\{
J_{a},P_{a},Z_{a},Q^{i}\right\} $ with $i=1,\dots ,\mathcal{N}$. In
particular the (anti)-commutation relations of the $\mathcal{N}=2$ centrally
extended Maxwell superalgebra are given by%
\begin{eqnarray}
\left[ J_{a},J_{b}\right] &=&\epsilon _{abc}J^{c}\,,\qquad \left[ J_{a},P_{b}%
\right] =\epsilon _{abc}P^{c}\,,  \notag \\
\left[ J_{a},Z_{b}\right] &=&\epsilon _{abc}Z^{c}\,,\qquad \left[ P_{a},P_{b}%
\right] =\epsilon _{abc}Z^{c}\,,  \notag \\
\left[ J_{a},Q_{\alpha }^{i}\right] &=&\frac{1}{2}\,\left( \Gamma
_{a}\right) _{\text{ }\alpha }^{\beta }Q_{\beta }^{i}\,,\text{ \ \ }\left[
J_{a},\Sigma _{\alpha }^{i}\right] =\frac{1}{2}\,\left( \Gamma _{a}\right) _{%
\text{ }\alpha }^{\beta }\Sigma _{\beta }^{i}\,,\text{ \ }  \label{2Ms} \\
\left[ P_{a},Q_{\alpha }^{i}\right] &=&\frac{1}{2}\,\left( \Gamma
_{a}\right) _{\text{ }\alpha }^{\beta }\Sigma _{\beta }^{i}\,,\text{ }
\notag \\
\left\{ Q_{\alpha }^{i},Q_{\beta }^{j}\right\} &=&-\frac{1}{2}\delta
^{ij}\left( C\Gamma ^{a}\right) _{\alpha \beta }P_{a}\,,  \notag \\
\left\{ Q_{\alpha }^{i},\Sigma _{\beta }^{j}\right\} &=&-\frac{1}{2}\,\delta
^{ij}\left( C\Gamma ^{a}\right) _{\alpha \beta }Z_{a}\,+C_{\alpha \beta
}\epsilon ^{ij}Z\,.  \notag
\end{eqnarray}%
Nevertheless, such central extension of the $\mathcal{N}=2$ Maxwell
superalgebra cannot reproduce a $\mathcal{N}=2$ CS supergravity action in
three spacetime dimensions. The $\left( 2,0\right) $ Maxwell superalgebra (%
\ref{2Ms}) does not have an invariant non-degenerate inner product. Indeed
the central charge $Z$ is orthogonal to the super Maxwell generators and to
itself. In order to have a non-degenerate invariant inner product, it is
necessary to enlarge the Maxwell superalgebra by introducing $\mathfrak{so}%
\left( 2\right) $ internal symmetry generators. In particular, we consider
two internal symmetry generators $T$ and $B$ such that they satisfy the
following non-trivial commutation relations:%
\begin{eqnarray}
\left[ Q_{\alpha }^{i},T\right] &=&\epsilon ^{ij}Q_{\alpha }^{j}\,,  \notag
\\
\left[ Q_{\alpha }^{i},B\right] &=&\epsilon ^{ij}\Sigma _{\alpha }^{j}\,, \\
\left[ \Sigma _{\alpha }^{i},T\right] &=&\epsilon ^{ij}\Sigma _{\alpha
}^{j}\,.  \notag
\end{eqnarray}%
Furthermore, one can see that the Jacobi identity requires that the
anticommutator of the Majorana spinor generators $Q^{i}$ has the following
form,%
\begin{equation}
\left\{ Q_{\alpha }^{i},Q_{\beta }^{j}\right\} =-\frac{1}{2}\delta
^{ij}\left( C\Gamma ^{a}\right) _{\alpha \beta }P_{a}+C_{\alpha \beta
}\epsilon ^{ij}B\,.
\end{equation}%
Remarkably, the $\mathcal{N}=2$ Maxwell superalgebra endowed with a central
charge $Z$ and two additional bosonic $\mathfrak{so}\left( 2\right) $
generators admits the following non-vanishing components of the invariant
tensor,%
\begin{eqnarray}
\left\langle J_{a}J_{b}\right\rangle &=&\alpha _{0}\eta _{ab}\,,\text{\qquad
}\left\langle P_{a}P_{b}\right\rangle =\alpha _{2}\eta _{ab}\,,  \notag \\
\left\langle J_{a}P_{b}\right\rangle &=&\alpha _{1}\eta _{ab}\,,\text{\qquad
}\left\langle Q_{\alpha }^{i}Q_{\beta }^{j}\right\rangle =\alpha
_{1}C_{\alpha \beta }\delta ^{ij}\,,  \notag \\
\left\langle J_{a}Z_{b}\right\rangle &=&\alpha _{2}\eta _{ab}\,,\text{\qquad
}\left\langle Q_{\alpha }^{i}\Sigma _{\beta }^{j}\right\rangle =\alpha
_{2}C_{\alpha \beta }\delta ^{ij}\,,  \label{InvTenMax2} \\
\left\langle TB\right\rangle &=&-\alpha _{1}\,,\qquad \ \ \ \left\langle
TZ\right\rangle =-\alpha _{2}\,,  \notag
\end{eqnarray}%
where $\alpha _{0},\alpha _{1}$ and $\alpha _{2}$ are real constants.
Well-defined invariant tensor for a $\mathcal{N}=2$ Maxwell superalgebra has
also been presented in \cite{CFR}. Nevertheless, the bosonic field content
is larger than our case since the $\mathcal{N}=2$ super Maxwell considered
in \cite{CFR} corresponds to a supersymmetric extension of a generalized
Maxwell algebra. Indeed, such generalization contains an extra bosonic gauge
field $\tilde{Z}_{ab}$ in addition to the usual Maxwell gauge fields.

The gauge connection one-form for the $\mathcal{N}=2$ super Maxwell reads%
\begin{equation}
A=\omega ^{a}J_{a}+e^{a}P_{a}+\sigma ^{a}Z_{a}+\bar{\psi}^{i}Q^{i}+\bar{\xi}%
^{i}\Sigma ^{i}+aT+bB+cZ\,,  \label{1f}
\end{equation}%
which, in addition to the usual Maxwell fields and $\mathcal{N}=2$
gravitini, contains three additional gauge fields given by $a,b$ and $c$,
respectively.

The curvature two form $F=dA+A^{2}$ is given by%
\begin{equation}
F=R^{a}J_{a}+\mathcal{T}^{a}P_{a}+\mathcal{F}^{a}Z_{a}+\nabla \bar{\psi}%
^{i}Q^{i}+\nabla \bar{\xi}^{i}\Sigma ^{i}\,+F\left( a\right) T+F\left(
b\right) B+F\left( c\right) Z\,,  \label{FS}
\end{equation}%
with%
\begin{eqnarray}
R^{a} &=&d\omega ^{a}+\frac{1}{2}\epsilon ^{abc}\omega _{b}\omega _{c}\,,
\notag \\
\mathcal{T}^{a} &=&de^{a}+\epsilon ^{abc}\omega _{b}e_{c}+\frac{1}{4}\bar{%
\psi}^{i}\Gamma ^{a}\psi ^{i}\,,  \notag \\
\mathcal{F}^{a} &=&d\sigma ^{a}+\epsilon ^{abc}\omega _{b}\sigma _{c}+\frac{1%
}{2}\epsilon ^{abc}e_{b}e_{c}+\frac{1}{2}\bar{\psi}^{i}\Gamma ^{a}\xi ^{i}\,,
\\
\nabla \psi ^{i} &=&d\psi ^{i}+\frac{1}{2}\,\omega ^{a}\Gamma _{a}\psi
^{i}+a\epsilon ^{ij}\psi ^{j}\,,  \notag \\
\nabla \xi ^{i} &=&d\xi ^{i}+\frac{1}{2}\,\omega ^{a}\Gamma _{a}\xi ^{i}\,+%
\frac{1}{2}\,e^{a}\Gamma _{a}\psi ^{i}+a\epsilon ^{ij}\xi ^{j}+b\epsilon
^{ij}\psi ^{j}\,,  \notag
\end{eqnarray}%
and%
\begin{eqnarray}
F\left( a\right) &=&\,da,  \notag \\
F\left( b\right) &=&db-\epsilon ^{ij}\psi _{i}\psi _{j}\,, \\
F\left( c\right) &=&dc-2\epsilon ^{ij}\psi _{i}\xi _{j}\,.  \notag
\end{eqnarray}

The CS supergravity form for the connection (\ref{1f}) constructed with the
invariant tensor (\ref{InvTenMax2}) defines a gauge-invariant supergravity
action for the $\mathcal{N}=2$ Maxwell superalgebra which is given by%
\begin{eqnarray}
I &=&\frac{k}{4\pi }\int \alpha _{0}\left( \,\omega ^{a}d\omega _{a}+\frac{1%
}{3}\,\epsilon _{abc}\omega ^{a}\omega ^{b}\omega ^{c}\right)   \notag \\
&&+\alpha _{1}\left( 2e^{a}R_{a}-\bar{\psi}^{i}\nabla \psi ^{i}-2adb\right)
\,+\alpha _{2}\left( 2R^{a}\sigma _{a}+e^{a}T_{a}-\bar{\psi}^{i}\nabla \xi
^{i}-\bar{\xi}^{i}\nabla \psi ^{i}-2adc\right) \,,  \label{SM2}
\end{eqnarray}%
up to a boundary term. Here, $T^{a}=de^{a}+\epsilon ^{abc}\omega _{b}e_{c}$
is the usual torsion two-form. Let us note that the term proportional to $%
\alpha _{1}$ coincides with the $\mathcal{N}=2$ Poincaré supergravity Lagrangian of \cite{HIPT}. The
Maxwell gravitational field $\sigma ^{a}$ and the additional Majorana spinor
field $\xi $ appear only in the exotic term proportional to $\alpha _{2}$.
Thus, the $\mathcal{N}=2$ supergravity action (\ref{SM2}) can be seen as a
Maxwell extension of the $\mathcal{N}=2$ Poincaré supergravity action
considered in \cite{HIPT}.

\section{$\mathcal{N}$-extended Maxwell Chern-Simons supergravity\qquad}

In this section we present the three-dimensional $\mathcal{N}$-extended
Maxwell supergravity theory based on the CS formulation. In particular, we
shall focus on the generic $\mathcal{N}=\left( \mathcal{N},0\right) $ case.
Nevertheless, our approach can be extended to the $\mathcal{N}=\left(
p,q\right) $ case.

A centrally $\mathcal{N}$-extended Maxwell superalgebra can be obtained by
considering $\mathcal{N}$ spinor generators $Q_{\alpha }^{i}$ and $\Sigma
_{\alpha }^{i}$,\ with $i=1,\dots ,\mathcal{N}$, in addition to the usual
Maxwell bosonic generators $\left\{ J_{a},P_{a},Z_{a}\right\} $ and the $%
\mathcal{N}\left( \mathcal{N}-1\right) /2$ central charges $Z^{ij}=-Z^{ji}$.
Then, an $\mathcal{N}$-extended supersymmetric central extension of the
Maxwell algebra is given by the following non-vanishing (anti-)commutation
relations%
\begin{eqnarray}
\left[ J_{a},J_{b}\right] &=&\epsilon _{abc}J^{c}\,,\qquad \left[ J_{a},P_{b}%
\right] =\epsilon _{abc}P^{c}\,,  \notag \\
\left[ J_{a},Z_{b}\right] &=&\epsilon _{abc}Z^{c}\,,\text{ \ \ \ \ }\left[
P_{a},P_{b}\right] =\epsilon _{abc}Z^{c}\,,  \notag \\
\left[ J_{a},Q_{\alpha }^{i}\right] &=&\frac{1}{2}\,\left( \Gamma
_{a}\right) _{\text{ }\alpha }^{\beta }Q_{\beta }^{i}\,,\text{ \ }  \notag \\
\text{\ }\left[ J_{a},\Sigma _{\alpha }^{i}\right] &=&\frac{1}{2}\,\left(
\Gamma _{a}\right) _{\text{ }\alpha }^{\beta }\Sigma _{\beta }^{i}\,,\text{
\ \ }  \label{SMN} \\
\left[ P_{a},Q_{\alpha }^{i}\right] &=&\frac{1}{2}\,\left( \Gamma
_{a}\right) _{\text{ }\alpha }^{\beta }\Sigma _{\beta }^{i}\,,\text{ }
\notag \\
\left\{ Q_{\alpha }^{i},Q_{\beta }^{j}\right\} &=&-\frac{1}{2}\delta
^{ij}\left( C\Gamma ^{a}\right) _{\alpha \beta }P_{a}\,\,,  \notag \\
\left\{ Q_{\alpha }^{i},\Sigma _{\beta }^{j}\right\} &=&-\frac{1}{2}\delta
^{ij}\,\left( C\Gamma ^{a}\right) _{\alpha \beta }Z_{a}\,+C_{\alpha \beta
}Z^{ij}\,.  \notag
\end{eqnarray}%
Although such supersymmetrization is well-defined, it does not allow to
construct a CS supergravity action. Indeed, the superalgebra (\ref{SMN})
does not have a non-degenerate invariant inner product. \ This can be seen
by noting that the $Z^{ij}$ generator is orthogonal to all generators of the
$\mathcal{N}$-extended Maxwell superalgebra and to itself. As in the $%
\mathcal{N}=2$ super Maxwell case, we have to enlarge the $\mathcal{N}$%
-extended Maxwell superalgebra by adding $\mathfrak{so}\left( N\right) $
generators. In particular, we have to include the new generators $%
T^{ij}=-T^{ji}$ and $B^{ij}=-B^{ji}$ which satisfy the following non-trivial
commutation relations:%
\begin{eqnarray}
\left[ T^{ij},T^{kl}\right] &=&\delta ^{jk}T^{il}-\delta ^{ik}T^{jl}-\delta
^{jl}T^{ik}+\delta ^{il}T^{jk}\,,  \notag \\
\left[ T^{ij},B^{kl}\right] &=&\delta ^{jk}B^{il}-\delta ^{ik}B^{jl}-\delta
^{jl}B^{ik}+\delta ^{il}B^{jk}\,,  \notag \\
\left[ B^{ij},B^{kl}\right] &=&\delta ^{jk}Z^{il}-\delta ^{ik}Z^{jl}-\delta
^{jl}Z^{ik}+\delta ^{il}Z^{jk}\,,  \notag \\
\left[ T^{ij},Z^{kl}\right] &=&\delta ^{jk}Z^{il}-\delta ^{ik}Z^{jl}-\delta
^{jl}Z^{ik}+\delta ^{il}Z^{jk}\,,  \label{SMN2} \\
\left[ T^{ij},Q_{\alpha }^{k}\right] &=&\left( \delta ^{jk}Q_{\alpha
}^{i}-\delta ^{ik}Q_{\alpha }^{j}\right) \,,  \notag \\
\left[ B^{ij},Q_{\alpha }^{k}\right] &=&\left( \delta ^{jk}\Sigma _{\alpha
}^{i}-\delta ^{ik}\Sigma _{\alpha }^{j}\right) \,,  \notag \\
\left[ T^{ij},\Sigma _{\alpha }^{k}\right] &=&\left( \delta ^{jk}\Sigma
_{\alpha }^{i}-\delta ^{ik}\Sigma _{\alpha }^{j}\right) \,.  \notag
\end{eqnarray}%
Additionally, the anticommutator of the Majorana spinors $Q_{\alpha }^{i}$
is modified as%
\begin{equation}
\left\{ Q_{\alpha }^{i},Q_{\beta }^{j}\right\} =-\frac{1}{2}\delta
^{ij}\left( C\Gamma ^{a}\right) _{\alpha \beta }P_{a}+C_{\alpha \beta
}B^{ij}\,.  \label{SMN3}
\end{equation}%
Let us note that such central $\mathcal{N}$-extension of the Maxwell
superalgebra endowed with $\mathfrak{so}\left( \mathcal{N}\right) $ internal
symmetry algebra is the three-dimensional version of the $D=4$ $\mathcal{N}$%
-extended Maxwell superalgebras obtained in \cite{AILW, CR1}, without
additional bosonic charges $\left\{ \bar{Z}_{\mu \nu },Z_{\mu }\right\} $. \
In particular, the present superalgebra can be seen as a deformation and
enlargement of the $\mathcal{N}$-extended Poincaré superalgebra discussed in
\cite{HIPT}. Interestingly, the $\mathcal{N}$-extended Maxwell superalgebra
with central charge and $\mathfrak{so}\left( \mathcal{N}\right) $ internal
symmetry algebra can be alternatively obtained as a semigroup expansion \cite%
{Sexp} of a $\mathcal{N}$-extended Lorentz superalgebra following the
procedure used in \cite{CPR}.

The central $\mathcal{N}$-extension of the Maxwell superalgebra endowed with
bosonic $\mathfrak{so}\left( \mathcal{N}\right) $ generators admits the
following non-vanishing components of the invariant tensor,%
\begin{eqnarray}
\left\langle J_{a}J_{b}\right\rangle &=&\alpha _{0}\eta _{ab}\,,\text{\qquad
}\left\langle P_{a}P_{b}\right\rangle =\alpha _{2}\eta _{ab}\,,  \notag \\
\left\langle J_{a}P_{b}\right\rangle &=&\alpha _{1}\eta _{ab}\,,\text{\qquad
}\left\langle Q_{\alpha }^{i}Q_{\beta }^{j}\right\rangle =\alpha
_{1}C_{\alpha \beta }\delta ^{ij}\,,  \notag \\
\left\langle J_{a}Z_{b}\right\rangle &=&\alpha _{2}\eta _{ab}\,,\text{\qquad
}\left\langle Q_{\alpha }^{i}\Sigma _{\beta }^{j}\right\rangle =\alpha
_{2}C_{\alpha \beta }\delta ^{ij}\,,  \notag \\
\left\langle T^{ij}T^{kl}\right\rangle &=&\alpha _{0}\left( \delta
^{ik}\delta ^{lj}-\delta ^{il}\delta ^{kj}\right) \,,  \label{itn} \\
\left\langle B^{ij}T^{kl}\right\rangle &=&\alpha _{1}\left( \delta
^{ik}\delta ^{lj}-\delta ^{il}\delta ^{kj}\right) \,,  \notag \\
\left\langle Z^{ij}T^{kl}\right\rangle &=&\alpha _{2}\left( \delta
^{ik}\delta ^{lj}-\delta ^{il}\delta ^{kj}\right) \,,  \notag \\
\left\langle B^{ij}B^{kl}\right\rangle &=&\alpha _{2}\left( \delta
^{ik}\delta ^{lj}-\delta ^{il}\delta ^{kj}\right) \,,  \notag
\end{eqnarray}%
where $\alpha _{0}$, $\alpha _{1}$ and $\alpha _{2}$ are real constants. Let
us note that the central charges $Z^{ij}$ are orthogonal to the generators
of the $\mathcal{N}$-extended Maxwell superalgebra. The presence of internal
symmetries generators allows to achieve appropriately a non-degenerate
invariant inner product.

The other crucial ingredient for the construction of a CS action is the
connection one-form which, for our case, reads%
\begin{equation}
A=\omega ^{a}J_{a}+e^{a}P_{a}+\sigma ^{a}Z_{a}+\bar{\psi}^{i}Q^{i}+\bar{\xi}%
^{i}\Sigma ^{i}+\frac{1}{2}a^{ij}T_{ij}+\frac{1}{2}b^{ij}B_{ij}+\frac{1}{2}%
c^{ij}Z_{ij}\,,  \label{c1fn}
\end{equation}%
where the coefficients in front of the generators are the gauge potential
one-forms. In particular, the theory contains $\mathcal{N}$ gravitini.

The corresponding curvature two-form is given by%
\begin{equation}
F=R^{a}J_{a}+\mathcal{T}^{a}P_{a}+\mathcal{F}^{a}Z_{a}+\nabla \bar{\psi}%
^{i}Q^{i}+\nabla \bar{\xi}^{i}\Sigma ^{i}+\frac{1}{2}F^{ij}\left( a\right)
T_{ij}+\frac{1}{2}F^{ij}\left( b\right) B_{ij}+\frac{1}{2}F^{ij}\left(
c\right) Z_{ij}\,,
\end{equation}%
where%
\begin{eqnarray}
R^{a} &=&d\omega ^{a}+\frac{1}{2}\epsilon ^{abc}\omega _{b}\omega _{c}\,,
\notag \\
\mathcal{T}^{a} &=&de^{a}+\epsilon ^{abc}\omega _{b}e_{c}+\frac{1}{4}\bar{%
\psi}^{i}\Gamma ^{a}\psi ^{i}\,,
\end{eqnarray}%
are the Lorentz and supertorsion curvature, respectively. On the other hand,
we have%
\begin{eqnarray}
\mathcal{F}^{a} &=&d\sigma ^{a}+\epsilon ^{abc}\omega _{b}\sigma _{c}+\frac{1%
}{2}\epsilon _{\text{ }}^{abc}e_{b}e_{c}+\frac{1}{2}\bar{\psi}^{i}\Gamma
^{a}\xi ^{i}\,,  \notag \\
\nabla \psi ^{i} &=&d\psi ^{i}+\frac{1}{2}\,\omega ^{a}\Gamma _{a}\psi
^{i}+a^{ij}\psi ^{j}\,,  \notag \\
\nabla \xi ^{i} &=&d\xi ^{i}+\frac{1}{2}\,\omega ^{a}\Gamma _{a}\xi ^{i}\,+%
\frac{1}{2}\,e^{a}\Gamma _{a}\psi ^{i}+a^{ij}\xi ^{j}+b^{ij}\,\psi ^{j},
\notag \\
F^{ij}\left( a\right) &=&da^{ij}+a^{ik}a^{kj}\,, \\
F^{ij}\left( b\right) &=&db^{ij}+a^{ik}b^{kj}+b^{ik}a^{kj}-\bar{\psi}%
^{i}\psi ^{j}\,,  \notag \\
F^{ij}\left( c\right) &=&dc^{ij}+a^{ik}c^{kj}+c^{ik}a^{kj}+b^{ik}b^{kj}-2%
\bar{\psi}^{i}\xi ^{j}\,.  \notag
\end{eqnarray}%
Then, using the non-vanishing components of the invariant tensor (\ref{itn})
and the connection one-form (\ref{c1fn}) in the explicit form of the CS
action (\ref{CSA}) we find,%
\begin{eqnarray}
I &=&\frac{k}{4\pi }\int \alpha _{0}\left[ \,\omega ^{a}d\omega _{a}+\frac{1%
}{3}\,\epsilon _{abc}\omega ^{a}\omega ^{b}\omega ^{c}-\frac{1}{2}\left(
a^{ij}da^{ji}+\frac{2}{3}a^{ik}a^{kj}a^{ji}\right) \right]  \notag \\
&&+\alpha _{1}\left[ 2e^{a}R_{a}-\bar{\psi}^{i}\nabla \psi
^{i}+b^{ji}F^{ij}\left( a\right) \right]  \label{SAn} \\
&&\,+\alpha _{2}\left[ 2R^{a}\sigma _{a}+e^{a}T_{a}-\bar{\psi}^{i}\nabla \xi
^{i}-\bar{\xi}^{i}\nabla \psi ^{i}+c^{ji}F^{ij}\left( a\right)
+b^{ji}f^{ij}\left( b\right) \right] \,,  \notag
\end{eqnarray}%
where%
\begin{equation}
f^{ij}\left( b\right) =db^{ij}+a^{ik}b^{kj}+b^{ik}a^{kj}\,.
\end{equation}%
It is interesting to note that the CS terms proportional to $\alpha _{0}$
and $\alpha _{1}$ is analogous to the centrally $\mathcal{N}$-extended
Poincaré supergravity Lagrangian. A subtle difference appear in the nature
of the $b^{ij}$ field which in the Poincaré case is related to the central
charge meanwhile here $b^{ij}$ is a $\mathfrak{so}\left( \mathcal{N}\right) $
gauge field. The gravitational Maxwell gauge field $\sigma _{a}$ and the
additional Majorana spinor charge $\xi ^{i}$ appear on the exotic sector $%
\alpha _{2}$. This is quite different to the minimal four-dimensional
Maxwell supergravity theory in which the extra fields $\sigma _{a}$ and $\xi
$ appear only on the boundary term without modifying the dynamics of flat
supergravity \cite{CRR}. As was discussed in \cite{CRR}, such extra field
were crucial to restore supersymmetry of the theory on a manifold with
boundary. Here, the supergravity action (\ref{SAn}) is invariant under the
following supersymmetry transformation laws%
\begin{eqnarray}
\delta \omega ^{a} &=&0\,,\qquad  \notag \\
\delta e^{a} &=&\frac{1}{2}\bar{\zeta}^{i}\Gamma ^{a}\psi ^{i}\,,  \notag \\
\delta \sigma ^{a} &=&\frac{1}{2}\bar{\zeta}^{i}\Gamma ^{a}\xi ^{i}+\frac{1}{%
2}\bar{\varrho}^{i}\Gamma ^{a}\psi ^{i}\,,  \notag \\
\delta \psi ^{i} &=&d\zeta ^{i}+\frac{1}{2}\omega ^{a}\Gamma _{a}\zeta
^{i}\,+a^{ij}\zeta ^{j}, \\
\delta \xi ^{i} &=&d\varrho ^{i}+\frac{1}{2}\omega ^{a}\Gamma _{a}\varrho
^{i}+a^{ij}\varrho ^{j}+\frac{1}{2}e^{a}\Gamma _{a}\zeta ^{i}+b^{ij}\zeta
^{j}\,  \notag \\
\delta a^{ij} &=&0\,,\qquad  \notag \\
\delta b^{ij} &=&-2\bar{\psi}^{\left[ i\right. }\zeta ^{\left. j\right] }\,,
\notag \\
\delta c^{ij} &=&-2\left( \bar{\psi}^{\left[ i\right. }\varrho ^{\left. j%
\right] }+\bar{\xi}^{\left[ i\right. }\zeta ^{\left. j\right] }\right) \,,
\notag
\end{eqnarray}%
where $\zeta ^{i}$ and $\varrho ^{i}$ are the fermionic gauge parameters
related to $Q^{i}$ and $\Sigma ^{i}$, respectively.

The equations of motion derived from the supergravity action (\ref{SAn})
reduce to the vanishing of the curvatures, for $\alpha _{2}\neq 0$,%
\begin{equation}
\begin{tabular}{lll}
$R^{a}=0\,,$ & $\mathcal{T}^{a}=0\,,$ & $\mathcal{F}^{a}=0\,,$ \\
$\nabla \psi ^{i}=0\,,$ & $\nabla \xi ^{i}=0\,,$ & $F^{ij}\left( a\right)
=0\,,$ \\
$F^{ij}\left( b\right) =0\,,$ & $F^{ij}\left( c\right) =0\,.$ &
\end{tabular}
\label{eom}
\end{equation}%
It is interesting to note that the vanishing of $\mathcal{F}^{a}$ is
analogue to the constancy of an abelian SUSY field strength background.
Although $\sigma ^{a}$ appears only on the exotic part, one could expect
that its presence would influence the boundary dynamics. In particular, one
could argue that the asymptotic symmetries of the $\mathcal{N}=2$ Maxwell
supergravity are spanned by a deformation and enlargement of the $\mathcal{N}%
=2$ super $\mathfrak{bms}_{3}$ algebra presented in \cite{FMT3, CCFR2} as
occurs in the bosonic sector \cite{CMMRSV, CCRS}.

The centrally $\mathcal{N}$-extended Maxwell supergravity enlarged with $%
\mathfrak{so}\left( \mathcal{N}\right) $ gauge fields allows us to define an
alternative supergravity model. It is interesting to note that we do not
introduce a cosmological constant term although we have enlarged the field
content. Similarly to the minimal case \cite{CPR}, one could recover our
result as a flat limit of a particular supergravity theory. Indeed, as we
shall see, one can add a length scale by modifying the superalgebra and the
supergravity action from which a Maxwell limit can be properly applied.

\section{$\mathcal{N}$-extended Maxwell supergravity theory as a flat limit}

The incorporation of a cosmological constant term in a $\mathcal{N}>2$
supergravity theory can be done by considering an extension of the $\left(
p,q\right) $ AdS supergravity by an $\mathfrak{so}\left( p\right) \oplus
\mathfrak{so}\left( q\right) $ automorphism algebra allowing a well-defined
flat limit \cite{HIPT}. Here we propose an alternative approach to
incorporate a cosmological constant to a $\mathcal{N} $-extended
supergravity by considering an enlarged superalgebra such that the flat
limit leads us to the $\mathcal{N}$-extended Maxwell theory.

A length parameter $\ell $ can be introduced to the set of generators $%
\left\{ J_{a},P_{a},Z_{a},Q^{i},\Sigma ^{i},T^{ij},B^{ij},Z^{ij}\right\} $
by considering a supersymmetric $\mathcal{N}$-extension of the AdS-Lorentz
symmetry. Such generators satisfy not only the non-vanishing
(anti-)commutators of the $\mathcal{N}$-extended Maxwell superalgebra (\ref%
{SMN})-(\ref{SMN3}) but also%
\begin{eqnarray}
\left[ Z_{a},Z_{b}\right] &=&\frac{1}{\ell ^{2}}\epsilon
_{abc}Z^{c}\,,\qquad \text{\ \ }  \notag \\
\left[ P_{a},Z_{b}\right] &=&\frac{1}{\ell ^{2}}\epsilon _{abc}P^{c}\,,
\notag \\
\left[ Z_{a},Q_{\alpha }^{i}\right] &=&\frac{1}{2\ell ^{2}}\,\left( \Gamma
_{a}\right) _{\text{ }\alpha }^{\beta }Q_{\beta }^{i}\,,\text{ }  \notag \\
\left[ P_{a},\Sigma _{\alpha }^{i}\right] &=&\frac{1}{2\ell ^{2}}\,\left(
\Gamma _{a}\right) _{\text{ }\alpha }^{\beta }Q_{\beta }^{i}\,,
\label{sadsl1} \\
\left[ Z_{a},\Sigma _{\alpha }^{i}\right] &=&\frac{1}{2\ell ^{2}}\,\left(
\Gamma _{a}\right) _{\text{ }\alpha }^{\beta }\Sigma _{\beta }^{i}\,,  \notag
\\
\left\{ \Sigma _{\alpha }^{i},\Sigma _{\beta }^{j}\right\} &=&-\frac{1}{%
2\ell ^{2}}\,\delta ^{ij}\left( C\Gamma ^{a}\right) _{\alpha \beta }P_{a}+%
\frac{1}{\ell ^{2}}C_{\alpha \beta }B^{ij}\,.  \notag
\end{eqnarray}%
\begin{eqnarray}
\left[ B^{ij},Z^{kl}\right] &=&\frac{1}{\ell ^{2}}\left( \delta
^{jk}B^{il}-\delta ^{ik}B^{jl}-\delta ^{jl}B^{ik}+\delta ^{il}ZB^{jk}\right)
\,,  \notag \\
\left[ Z^{ij},Z^{kl}\right] &=&\frac{1}{\ell ^{2}}\left( \delta
^{jk}Z^{il}-\delta ^{ik}Z^{jl}-\delta ^{jl}Z^{ik}+\delta ^{il}Z^{jk}\right)
\,,  \notag \\
\left[ B^{ij},\Sigma _{\alpha }^{k}\right] &=&\frac{1}{\ell ^{2}}\left(
\delta ^{jk}Q_{\alpha }^{i}-\delta ^{ik}Q_{\alpha }^{j}\right) \,,
\label{sadsl2} \\
\left[ Z^{ij},Q_{\alpha }^{k}\right] &=&\frac{1}{\ell ^{2}}\left( \delta
^{jk}Q_{\alpha }^{i}-\delta ^{ik}Q_{\alpha }^{j}\right) \,,  \notag \\
\left[ Z^{ij},\Sigma _{\alpha }^{k}\right] &=&\frac{1}{\ell ^{2}}\left(
\delta ^{jk}\Sigma _{\alpha }^{i}-\delta ^{ik}\Sigma _{\alpha }^{j}\right)
\,.  \notag
\end{eqnarray}

The superalgebra given by (\ref{SMN})-(\ref{SMN3}) and (\ref{sadsl1})-(\ref%
{sadsl2}) corresponds to a supersymmetric extension of the so called
AdS-Lorentz algebra enlarged with $\mathfrak{so}\left( \mathcal{N}\right) $
generators. Although its supersymmetrization is not unique and have been
explored with different purposes \cite{FISV, CRS, CIRR, BR}, this is the
smallest one with $2\mathcal{N}$ spinor charges. Let us note that the $%
\mathfrak{so}\left( \mathcal{N}\right) $ generators, which satisfy (\ref%
{sadsl2}), are required in order to relate the $\mathcal{N}$-extended
AdS-Lorentz superalgebra with the centrally $\mathcal{N}$-extended Maxwell
algebra endowed with $\mathfrak{so}\left( \mathcal{N}\right) $ internal
symmetry. Naurally, the Maxwell limit $\ell \rightarrow \infty $ can be
applied properly leading to the $\mathcal{N}$-extended Maxwell superalgebra
with central charge and $\mathfrak{so}\left( \mathcal{N}\right) $ algebra.
It is interesting to point out that the $Z^{ij}$ generator becomes a central
charge after the flat limit. As we shall see, such enlargement allows us to
relate the non-degenerate invariant inner product to the Maxwell ones
through the Maxwell limit.

The $\mathcal{N}$-extended AdS-Lorentz superalgebra with $\mathfrak{so}%
\left( \mathcal{N}\right) $ generators admits the following non-vanishing
components of the invariant tensor,%
\begin{eqnarray}
\left\langle J_{a}J_{b}\right\rangle &=&\alpha _{0}\eta _{ab}\,,\text{\qquad
}\left\langle P_{a}P_{b}\right\rangle =\alpha _{2}\eta _{ab}\,,\qquad
\left\langle Q_{\alpha }^{i}Q_{\beta }^{j}\right\rangle =\alpha
_{1}C_{\alpha \beta }\delta ^{ij}\,,  \notag \\
\left\langle J_{a}P_{b}\right\rangle &=&\alpha _{1}\eta _{ab}\,,\text{\qquad
}\left\langle P_{a}Z_{b}\right\rangle =\frac{\alpha _{1}}{\ell ^{2}}\eta
_{ab}\,,\qquad \left\langle Q_{\alpha }^{i}\Sigma _{\beta }^{j}\right\rangle
=\alpha _{2}C_{\alpha \beta }\delta ^{ij}\,,  \label{ital} \\
\left\langle J_{a}Z_{b}\right\rangle &=&\alpha _{2}\eta _{ab}\,,\text{\qquad
}\left\langle Z_{a}Z_{b}\right\rangle =\frac{\alpha _{2}}{\ell ^{2}}\eta
_{ab}\,,\qquad \left\langle \Sigma _{\alpha }^{i}\Sigma _{\beta
}^{j}\right\rangle =\frac{\alpha _{1}}{\ell ^{2}}C_{\alpha \beta }\delta
^{ij}\,,  \notag
\end{eqnarray}%
\begin{eqnarray}
\left\langle T^{ij}T^{kl}\right\rangle &=&\alpha _{0}\left( \delta
^{ik}\delta ^{lj}-\delta ^{il}\delta ^{kj}\right) \,,\quad \left\langle
B^{ij}T^{kl}\right\rangle =\alpha _{1}\left( \delta ^{ik}\delta ^{lj}-\delta
^{il}\delta ^{kj}\right) \,,  \notag \\
\left\langle Z^{ij}T^{kl}\right\rangle &=&\alpha _{2}\left( \delta
^{ik}\delta ^{lj}-\delta ^{il}\delta ^{kj}\right) \,,\quad \left\langle
B^{ij}B^{kl}\right\rangle =\alpha _{2}\left( \delta ^{ik}\delta ^{lj}-\delta
^{il}\delta ^{kj}\right) \,,  \label{ital2} \\
\left\langle B^{ij}Z^{kl}\right\rangle &=&\frac{\alpha _{1}}{\ell ^{2}}%
\left( \delta ^{ik}\delta ^{lj}-\delta ^{il}\delta ^{kj}\right) \,,\quad
\left\langle Z^{ij}Z^{kl}\right\rangle =\frac{\alpha _{2}}{\ell ^{2}}\left(
\delta ^{ik}\delta ^{lj}-\delta ^{il}\delta ^{kj}\right) \,.  \notag
\end{eqnarray}%
One can see that the limit $\ell \rightarrow \infty $ reproduces the
invariant tensor of the $\mathcal{N}$-extended Maxwell superalgebra with
central charges and $\mathfrak{so}\left( \mathcal{N}\right) $ generators.

Although the connection one-form $A$ is analogous to the Maxwell one,%
\begin{equation}
A=\omega ^{a}J_{a}+e^{a}P_{a}+\sigma ^{a}Z_{a}+\bar{\psi}^{i}Q^{i}+\bar{\xi}%
^{i}\Sigma ^{i}+\frac{1}{2}a^{ij}T_{ij}+\frac{1}{2}b^{ij}B_{ij}+\frac{1}{2}%
c^{ij}Z_{ij}\,,  \label{1fadsl}
\end{equation}%
the corresponding curvature two-form is subtle different due to the new
commutators. Indeed, we have%
\begin{equation}
F=R^{a}J_{a}+\mathcal{T}^{a}P_{a}+\mathcal{F}^{a}Z_{a}+\nabla \bar{\psi}%
^{i}Q^{i}+\nabla \bar{\xi}^{i}\Sigma ^{i}+\frac{1}{2}F^{ij}\left( a\right)
T_{ij}+\frac{1}{2}F^{ij}\left( b\right) B_{ij}+\frac{1}{2}F^{ij}\left(
c\right) Z_{ij}\,,
\end{equation}%
where%
\begin{eqnarray}
R^{a} &=&d\omega ^{a}+\frac{1}{2}\epsilon ^{abc}\omega _{b}\omega _{c}\,,
\notag \\
\mathcal{T}^{a} &=&de^{a}+\epsilon ^{abc}\omega _{b}e_{c}+\frac{1}{\ell ^{2}}%
\epsilon ^{abc}\sigma _{b}e_{c}+\frac{1}{4}\bar{\psi}^{i}\Gamma ^{a}\psi
^{i}\,,  \notag \\
\mathcal{F}^{a} &=&d\sigma ^{a}+\epsilon ^{abc}\omega _{b}\sigma _{c}+\frac{1%
}{2\ell ^{2}}\epsilon ^{abc}\sigma _{b}\sigma _{c}+\frac{1}{2}\epsilon _{%
\text{ }}^{abc}e_{b}e_{c}+\frac{1}{2}\bar{\psi}^{i}\Gamma ^{a}\xi ^{i}\,,
\notag \\
F^{ij}\left( a\right) &=&da^{ij}+a^{ik}a^{kj}\,,  \label{ctf} \\
F^{ij}\left( b\right) &=&db^{ij}+a^{ik}b^{kj}+b^{ik}a^{kj}+\frac{1}{\ell ^{2}%
}\left( b^{ik}c^{kj}+c^{ik}b^{kj}\right) -\bar{\psi}^{i}\psi ^{j}-\frac{1}{%
\ell ^{2}}\bar{\xi}^{i}\xi ^{j}\,,  \notag \\
F^{ij}\left( c\right) &=&dc^{ij}+a^{ik}c^{kj}+c^{ik}a^{kj}+b^{ik}b^{kj}+%
\frac{1}{\ell ^{2}}c^{ik}c^{kj}-2\bar{\psi}^{i}\xi ^{j}\,.  \notag
\end{eqnarray}%
Here the covariant derivative acting on spinors read%
\begin{eqnarray}
\nabla \psi ^{i} &=&d\psi ^{i}+\frac{1}{2}\,\omega ^{a}\Gamma _{a}\psi
^{i}+a^{ij}\psi ^{j}+\frac{1}{\ell ^{2}}\left( b^{ij}\xi ^{j}+c^{ij}\psi
^{j}\right) \,,  \notag \\
\nabla \xi ^{i} &=&d\xi ^{i}+\frac{1}{2}\,\omega ^{a}\Gamma _{a}\xi ^{i}\,+%
\frac{1}{2}\,e^{a}\Gamma _{a}\psi ^{i}+a^{ij}\xi ^{j}+b^{ij}\,\psi ^{j}+%
\frac{1}{\ell ^{2}}c^{ij}\xi ^{j}\,.
\end{eqnarray}

The CS supergravity action based on the $\mathcal{N}$-extended AdS-Lorentz
superalgebra with $\mathfrak{so}\left( \mathcal{N}\right) $ generators and
the invariant tensor (\ref{ital})-(\ref{ital2}) reads, up to boundary terms%
\begin{eqnarray}
I &=&\frac{k}{4\pi }\int \alpha _{0}\left[ \,\omega ^{a}d\omega _{a}+\frac{1%
}{3}\,\epsilon _{abc}\omega ^{a}\omega ^{b}\omega ^{c}-\frac{1}{2}\left(
a^{ij}da^{ji}+\frac{2}{3}a^{ik}a^{kj}a^{ji}\right) \right]  \notag \\
&&+\alpha _{1}\left[ 2e^{a}R_{a}+\frac{1}{3\ell ^{2}}e^{a}e^{b}e^{c}+\frac{2%
}{\ell ^{2}}e^{a}F_{a}-\bar{\psi}^{i}\nabla \psi ^{i}-\frac{1}{\ell ^{2}}%
\bar{\xi}^{i}\nabla \xi ^{i}+b^{ji}F^{ij}\left( a\right) +\frac{1}{\ell ^{2}}%
b^{ji}f^{ij}\left( c\right) \right]  \notag \\
&&\,+\alpha _{2}\left[ 2R^{a}\sigma _{a}+e^{a}T_{a}+\frac{2}{\ell ^{2}}%
F^{a}\sigma _{a}+\frac{1}{\ell ^{2}}\epsilon _{abc}e^{a}\sigma ^{b}e^{c}-%
\bar{\psi}^{i}\nabla \xi ^{i}-\bar{\xi}^{i}\nabla \psi ^{i}\right. \\
&&\left. +c^{ji}F^{ij}\left( a\right) +b^{ji}f^{ij}\left( b\right) +\frac{1}{%
\ell ^{2}}c^{ji}f^{ij}\left( c\right) \right] \,,  \notag
\end{eqnarray}%
where we have defined%
\begin{eqnarray}
F^{a} &=&d\sigma ^{a}+\epsilon ^{abc}\omega _{b}\sigma _{c}+\frac{1}{2\ell
^{2}}\epsilon ^{abc}\sigma _{b}\sigma _{c}\,,  \notag \\
f^{ij}\left( b\right) &=&db^{ij}+a^{ik}b^{kj}+b^{ik}a^{kj}+\frac{1}{\ell ^{2}%
}\left( b^{ik}c^{kj}+c^{ik}b^{kj}\right) \,, \\
f^{ij}\left( c\right) &=&dc^{ij}+a^{ik}c^{kj}+c^{ik}a^{kj}+b^{ik}b^{kj}+%
\frac{1}{\ell ^{2}}c^{ik}c^{kj}\,.  \notag
\end{eqnarray}%
The $\mathcal{N}$-extended AdS-Lorentz symmetry offer us an alternative
procedure to introduce a cosmological constant term to a three-dimensional
CS supergravity action. In particular, the inclusion of $\mathfrak{so}\left(
\mathcal{N}\right) $ gauge fields allows to establish a well-defined flat
limit $\ell \rightarrow \infty $ leading to the central $\mathcal{N}$%
-extension Maxwell supergravity action enlarged with $\mathfrak{so}\left(
\mathcal{N}\right) $ CS terms. A particular difference with the Maxwell
theory is the explicit presence of the $\sigma ^{a}$ and $\xi ^{i}$ field on
the term proportional to $\alpha _{1}$. Such presence allows us to recover
the vanishing of the AdS-Lorentz curvature two forms (\ref{ctf}) as field
equations when $\alpha _{2}\neq 0$. Naturally, the Maxwell limit applied to
the equations of motions reproduces the field equations (\ref{eom}) of the $%
\mathcal{N}$-extended Maxwell supergravity theory with central charges and $%
\mathfrak{so}\left( \mathcal{N}\right) $ gauge fields.

\section{Discussion}

We have presented a new class of three-dimensional $\mathcal{N}$-extended CS
supergravity theories based on the centrally $\mathcal{N}$-extended Maxwell
superalgebra enlarged with $\mathfrak{so}(\mathcal{N})$ internal symmetry
algebra. The construction of the supergravity action requires to introduce $%
\mathfrak{so}(\mathcal{N})$ generators which are essential to the obtention
of non-degenerate invariant inner product. The new theories correspond to a $%
\mathcal{N}$-extension of the minimal Maxwell supergravity theory presented
in \cite{CPR} and can be seen as alternative $\mathcal{N}$-extended
supergravity theories without cosmological constant. Let us note that the CS
supergravity action is characterized by three coupling constants $\alpha
_{0} $, $\alpha _{1}$ and $\alpha _{2}$. Interestingly, the term
proportional to $\alpha _{0}$ and $\alpha _{1}$ is the usual $\mathcal{N}$%
-extended Poincaré action with $\mathfrak{so}\left( \mathcal{N}\right) $
gauge fields. On the other hand, the gravitational Maxwell field $\sigma
_{a} $ and the additional Majorana spinor field $\xi $ appear only on the $%
\alpha _{2}$ sector.

The introduction of a cosmological constant term to our model is achieved by
considering the $\mathcal{N}$-extended AdS-Lorentz superalgebra enlarged
with $\mathfrak{so}(\mathcal{N})$ generators. The presence of the $\mathfrak{%
so}(\mathcal{N})$ generators allows us to have a well-defined Maxwell limit $%
\ell \rightarrow \infty $ in which the $\mathcal{N}$-extended AdS-Lorentz
theory reduces properly to the $\mathcal{N}$-extended Maxwell one.

A proper characteristic of the Maxwell symmetries is the presence of the
gravitational Maxwell gauge field $\sigma _{a}$ which is related to the
abelian generator $Z_{a}$. Such generator modifies the asymptotic $\mathfrak{%
bms}_{3}$ symmetry of standard Einstein gravity to a deformed and enlarged $%
\mathfrak{bms}_{3}$ symmetry \cite{CMMRSV}. It would then be interesting to
extend the results of \cite{CMMRSV} to the minimal Maxwell supergravity
presented in \cite{CPR} and to the $\mathcal{N}$-extended Maxwell
supergravity theories presented here [work in progress].

On the other hand, it was recently shown that the deformed and enlarged $%
\mathfrak{bms}_{3}$ algebra can be obtained as a flat limit of three copies
of the Virasoro algebra \cite{CCRS}. Such asymptotic symmetry results to be
the asymptotic structure of the AdS-Lorentz CS gravity \cite{CMRSV}. It
would be interesting to study these relations at the supersymmetric level.

\section{Acknowledgment}

This work was supported by the Chilean FONDECYT Projects N$^{\circ }$%
3170437. The author would like to thank to Octavio Fierro and Evelyn Rodrí%
guez for valuable discussions and comments.


\bigskip

\begin{thebibliography}{99}
\bibitem{DK} S. Deser, J.H. Kay, \textit{Topologically massive supergravity}%
, Phys. Lett. B \textbf{120} (1983) 97.

\bibitem{Deser} S. Deser, \textit{Cosmological Topological Supergravity,
Quantum Theory of Gravity: Essays in honor of the 60th Birthday of Bryce S},
(DeWitt. Published by Adam Hilger Ltd.,Bristol, 1984).

\bibitem{Nieuwenhuizen} P. van Nieuwenhuizen, \textit{Three-dimensional
conformal supergravity and Chern-Simons terms}, Phys. Rev. D \textbf{32}
(1985) 872.

\bibitem{AT} A. Achucarro, P.K. Townsend, \textit{A Chern-Simons action for
three-dimensional anti-De Sitter supergravity theories}, Phys. Lett. B
\textbf{180} (1986) 89.

\bibitem{RN} M. Rocek, P. van Nieuwenhuizen,\textit{\ }$N\geq 2$\textit{\
supersymmetric Chern-Simons terms as }$d=3$\textit{\ extended conformal
supergravity}, Class. Quant. Grav. \textbf{3} (1986) 43.

\bibitem{Witten} E. Witten, \textit{(2+1)-Dimensional gravity as an exactly
soluble system}, Nucl. Phys. B \textbf{311} (1988) 46.

\bibitem{AT2} A. Achucarro, P.K. Townsend, \textit{Extended supergravities
in }$d=\left( 2+1\right) $ \textit{as chern-simons theories}, Phys. Lett. B
\textbf{229} (1989) 383.

\bibitem{NG} H. Nishino, S.J. Gates Jr., \textit{Chern-Simons theories with
supersymmetries in three-dimensions}, Mod. Phys. A \textbf{8} (1993) 3371.

\bibitem{HIPT} P.S. Howe, J.M. Izquierdo, G. Papadopoulos, P.K. Townsend,
\textit{New supergravities with central charges and Killing spinors in 2+1
dimensions}, Nucl. Phys. B \textbf{467} (1996) 183. [hep-th/9505032].

\bibitem{BTZ} M. Banados, R. Troncoso, J. Zanelli, \textit{Higher
dimensional Chern-Simons supergravity}, Phys. Rev. D \textbf{54} (1996)
2605. [gr-qc/9601003].

\bibitem{GTW} A. Giacomini, R. Troncoso, S. Willison, \textit{%
Three-dimensional supergravity reloaded}, Class. Quant. Grav. \textbf{24}
(2007) 2845. [hep-th/0610077].

\bibitem{Sorokas} D.V. Soroka, V.A. Soroka, \textit{Tensor extension of the
Poincaré algebra}, Phys. Lett. B \textbf{607} (2005) 302. [hep-th/0410012].

\bibitem{SS} D.V. Soroka, V.A. Soroka,\textit{\ Semi-simple extension of the
(super) Poincaré algebra, }Adv. High Energy Phys. \textbf{2009} (2009)
234147. [hep-th/0605251].

\bibitem{DKGS} R. Durka, J. Kowalski-Glikman, M. Szczachor, \textit{Gauged
AdS-Maxwell algebra and gravity}, Mod. Phys. Lett. A \textbf{26} (2011)
2689. arXiv:1107.4728 [hep-th].

\bibitem{DFIMRSV} J. Díaz, O. Fierro, F. Izaurieta, N. Merino, E. Rodriguez,
P. Salgado, O. Valdivia, \textit{A generalized action for (2+1)-dimensional
Chern--Simons gravity, }J. Phys. A. Math. Theor. \textbf{45} (2012) 255207.

\bibitem{SalSal} P. Salgado, S. Salgado, $\mathfrak{so}\left( D-1,1\right)
\oplus \mathfrak{so}\left( D-1,2\right) $ \textit{algebras and gravity},
Phys. Lett. B \textbf{728} (2014) 5.

\bibitem{GRSS} N. González, G. Rubio, P. Salgado, S. Salgado, \textit{%
Einstein-Hilbert action with cosmological term from Chern-Simons gravity},
J. Geom. Phys. \textbf{86} (2014) 339. arXiv:1605.00325 [math-ph].

\bibitem{FMT} O. Fuentealba, J. Matulich, R. Troncoso, \textit{Extension of
the Poincaré group with half-integer spin generators: hypergravity and beyond%
}, JHEP \textbf{1509} (2015) 003. arXiv:1505.06173 [hep-th].

\bibitem{FMT2} O. Fuentealba, J. Matulich, R. Troncoso, \textit{%
Asymptotically flat structure of hypergravity in three spacetime dimensions}%
, JHEP \textbf{1510} (2015) 009. arXiv:1508.04663 [hep-th].

\bibitem{CDMR} P.K. Concha, R. Durka, N. Merino, E.K. Rodríguez, \textit{New
family of Maxwell like algebras}, Phys. Lett. B \textbf{759} (2016) 507.
arXiv:1601.06443 [hep-th].

\bibitem{Durka} R. Durka, \textit{Resonant algebras and gravity}, J. Phys. A%
\textbf{50} (2017) 145202. arXiv:1605.00059 [hep-th].

\bibitem{BDR} R. Basu, S. Detournay, M. Riegler \textit{Spectral Flow in 3D
Flat Spacetimes}, JHEP \textbf{1712} (2017) 134. arXiv:1706.07438 [hep-th].

\bibitem{CCFRS} R. Caroca, P. Concha, O. Fierro, E. Rodríguez, P.
Salgado-Rebolledo, \textit{Generalized Chern-Simons higher-spin gravity
theories in three dimensions}, Nucl. Phys. B \textbf{934} (2018) 240.
arXiv:1712.09975 [hep-th].

\bibitem{FISV} O. Fierro, F. Izaurieta, P. Salgado, O. Valdivia, \textit{%
Minimal AdS-Lorentz supergravity in three-dimensions}, Phys. Lett. B \textbf{%
788} (2019) 198. arXiv:1401.3697 [hep-th].

\bibitem{CMRSV} P. Concha, N. Merino, E. Rodríguez, P. Salgado-Rebolledo, O.
Valdivia, Semi-simple enlargement of the $\mathfrak{bms}_{3}$ \textit{%
algebra from a }$\mathfrak{so}\left( 2,2\right) \oplus \mathfrak{so}\left(
2,1\right) $ \textit{Chern-Simons theory}, JHEP \textbf{1902} (2019) 002.
arXiv:1810.12256 [hep-th].

\bibitem{Valcarcel} C.E. Valcárcel, \textit{New boundary conditions for
(extended) $AdS_3$ supergravity}, Class. Quant. Grav. \textbf{36} (2019)
065002. arXiv:1812.02799 [hep-th].

\bibitem{CPR} P. Concha, D.M. Peñafiel, E. Rodríguez, \textit{On the Maxwell
supergravity and flat limit in 2+1 dimensions}, Phys. Lett. B \textbf{785}
(2018) 247. arXiv:1807.00194 [hep-th].

\bibitem{BCR} H. Bacry, P. Combe, J.L. Richard, \textit{Group-theoretical
analysis of elementary particles in an external electromagnetic fields. 1.
The relativistic particle in a constant and uniform field}, Nuovo Cim. A
\textbf{67} (1970) 267.

\bibitem{Schrader} R. Schrader, \textit{The Maxwell group and the quantum
theory of particles in classical homogeneous electromagnetic fields},
Fortsch. Phys. \textbf{20} (1972) 701.

\bibitem{GK} J. Gomis, A. Kleinschmidt, \textit{On free Lie algebras and
particles in electro-magnetic fields}, JHEP \textbf{07} (2017) 085.
arXiv:1705.05854 [hep-th].

\bibitem{CPRS1} P.K. Concha, D.M. Peñafiel, E.K. Rodríguez, P. Salgado,
\textit{Even-dimensional General Relativity from Born-Infeld gravity}, Phys.
Lett. B \textbf{725} (2013) 419. arXiv:1309.0062 [hep-th].

\bibitem{CPRS2} P.K. Concha, D.M. Peñafiel, E.K. Rodríguez, P. Salgado,
\textit{Chern-Simons and Born-Infeld gravity theories and Maxwell algebras
type}, Eur. Phys. J. C \textbf{74} (2014) 2741. arXiv:1402.0023 [hep-th].

\bibitem{CPRS3} P.K. Concha, D.M. Peñafiel, E.K. Rodríguez, P. Salgado,
\textit{Generalized Poincaré algebras and Lovelock-Cartan gravity theory},
Phys. Lett. B \textbf{742} (2015) 310. arXiv:1405.7078 [hep.th].

\bibitem{SSV} P. Salgado, R.J. Szabo, O. Valdivia, \textit{Topological
gravity and transgression holography}, Phys. Rev. D\textbf{89} (2014)
084077. arXiv:1401.3653 [hep-th].

\bibitem{HR} S. Hoseinzadeh, A. Rezaei-Aghdam, \textit{(2+1)-dimensional
gravity from Maxwell and semisimple extension of the Poincaré gauge
symmetric models}, Phys. Rev. D\textbf{90} (2014) 084008. arXiv:1402.0320
[hep-th].

\bibitem{AFGHZ} L. Avilés, E. Frodden, J. Gomis, D. Hidalgo, J. Zanelli,
\textit{Non-Relativistic Maxwell Chern-Simons Gravity}, JHEP \textbf{1805 }%
(2018) 047. arXiv:1802.08453 [hep-th].

\bibitem{CMMRSV} P. Concha, N. Merino, O. Miskovic, E. Rodríguez, P.
Salgado-Rebolledo, O. Valdivia, \textit{Asymptotic symmetries of
three-dimensional Chern-Simons gravity for the Maxwell algebra}, JHEP
\textbf{1810} (2018) 079. arXiv:1805.08834 [hep-th].

\bibitem{GKL} J. Gomis, K. Kamimura, J. Lukierski, \textit{Deformations of
Maxwell algebra and their Dynamical Realizations}, JHEP \textbf{0908} (2009)
039. arXiv:0906.4464 [hep-th].

\bibitem{CDIMR} P.K. Concha, R. Durka, C. Inostroza, N. Merino, E.K. Rodrí%
guez, \textit{Pure Lovelock gravity and Chern-Simons theory}, Phys. Rev. D
\textbf{94} (2016) 024055. arXiv:1603.09424 [hep-th],

\bibitem{CMR} P.K. Concha, N. Merino, E.K. Rodríguez, \textit{Lovelock
gravity from Born-Infeld gravity theory}, Phys. Lett. B \textbf{765} (2017)
395. arXiv:1606.07083 [hep-th].

\bibitem{CR3} P. Concha, E. Rodríguez, \textit{Generalized Pure Lovelock
Gravity}, Phys. Lett. B \textbf{774} (2017) 616. arXiv:1708.08827 [hep-th].

\bibitem{BGKL} S. Bonanos, J. Gomis, K. Kamimura, J. Lukierski, \textit{%
Maxwell superalgebra and superparticle in constant Gauge background}. Phys.
Rev. Lett. \textbf{104} (2010) 090401. arXiv:0911.5072 [hep-th].

\bibitem{AI} J.A. de Azcarraga, J.M. Izquierdo, \textit{Minimal D=4
supergravity from superMaxwell algebra}, Nucl. Phys. B \textbf{885} (2014)
34. arXiv:1403.4128 [hep-th].

\bibitem{CR2} P.K. Concha, E.K. Rodríguez, \textit{N=1 Supergravity and
Maxwell superalgebras}, JHEP \textbf{1409} (2014) 090. arXiv:1407.4635
[hep-th].

\bibitem{CRR} P. Concha, L. Ravera, E. Rodríguez, \textit{On the
supersymmetry invariance of flat supergravity with boundary}, JHEP \textbf{01%
} (2019) 192. arXiv:1809.07871 [hep-th].

\bibitem{GKL2} J. Gomis, K. Kamimura, J. Lukierski, \textit{Deformations of
Maxwell algebras and their Dynamical Realizations}, JHEP \textbf{0908}
(2009) 039. arXiv:0906.4464 [hep-th].

\bibitem{BGKL2} S. Bonanos, J. Gomis, K. Kamimura, J. Lukierski, \textit{%
Deformations of Maxwell Superalgebras and Their Applications}, J. Math.
Phys. \textbf{51} (2010) 102301. arXiv:1005.3714 [hep-th].

\bibitem{Lukierski} J. Lukierski, \textit{Generalized Wigner-Inönü
Contractions and Maxwell (Super)Algebras}, Proc. Steklov Inst. Math. \textbf{%
272} (2011) no.1 183. arXiv:1007.3405 [hep-th].

\bibitem{FL} S. Fedoruk, J. Lukierski, \textit{New spinorial particle model
in tensorial space-time and interacting higher spin fields}, JHEP \textbf{%
1302} (2013) 128. arXiv:1210.1506 [hep-th].

\bibitem{CK} O. Cebecio\u{g}lu, S. Kibaro\u{g}lu,\textit{\ Maxwell-affine
gauge theory of gravity}, Phys. Lett. B \textbf{751} (2015) 131.
arXiv:1503.09003 [hep-th].

\bibitem{CFRS} P.K. Concha, O. Fierro, E.K. Rodríguez, P. Salgado, \textit{%
Chern-Simons supergravity in D=3 and Maxwell superalgebra}, Phys. Lett. B
\textbf{750} (2015) 117. arXiv:1507.02335 [hep-th].

\bibitem{CFR} P.K. Concha, O. Fierro, E.K. Rodríguez, \textit{Inönü-Wigner
contraction and D=2+1 supergravity}, Eur. Phys. J. C \textbf{77} (2017) 48.
arXiv:1611.05018 [hep-th].

\bibitem{PR} D.M. Peñafiel, L. Ravera, \textit{On the Hidden Maxwell
Superalgebra underlying D=4 Supergravity}, Fortsch. Phys. \textbf{65} (2017)
1700005. arXiv:1701.04234 [hep-th]

\bibitem{Ravera} L. Ravera, \textit{Hidden role of Maxwell superalgebras in
the free differential algebras of }$D=4$\textit{\ and }$D=11$\textit{\
supergravity}, Eur. Phys. J. C \textbf{78} (2018) 211. arXiv:1801.08860
[hep-th].

\bibitem{KSC} S. Kibaro\u{g}lu, M. \c{S}enay, O. Cebecio\u{g}lu, $D=4\mathit{%
\ }$\textit{topological gravity from gauging the Maxwell-special-affine group%
}, Mod. Phys. Lett. A\textbf{34} (2019) 1950016. arXiv:1810.01635 [hep-th].

\bibitem{KC} S. Kibaro\u{g}lu, O. Cebecio\u{g}lu, $D=4$\textit{\
supergravity from the Maxwell-Weyl superalgebra}, arXiv:1812.09861 [hep-th].

\bibitem{AF} R. D'Auria, P. Fré, \textit{Geometric supergravity in d=11 and
its hidden supergroup}, Nucl. Phys. B \textbf{201} (1982) 101.

\bibitem{Green} M.B. Green, \textit{Supertranslations, superstrings and
Chern-Simons forms}. Phys. Lett. B \textbf{223} (1989) 157.

\bibitem{AILW} J.A. de Azcarraga, J.M. Izquierdo, J. Lukierski, M.
Woronowicz, \textit{Generalizations of Maxwell (super)algebras by the
expansion method}, Nucl. Phys. B \textbf{869} (2013) 303. arXiv:1210.1117
[hep-th].

\bibitem{CR1} P.K. Concha, E.K. Rodríguez, \textit{Maxwell superalgebras and
Abelian semigroup expansion}, Nucl. Phys. B \textbf{886} (2014) 1128.
arXiv:1405.1334 [hep-th].

\bibitem{Sexp} F. Izaurieta, E. Rodríguez, P. Salgado, \textit{Expanding Lie
(super)algebras through Abelian semigroups}, J. Math. Phys. \textbf{47}
(2006) 123512. [hep-th/0606215].

\bibitem{FMT3} O. Fuentealba, J. Matulich, R. Troncoso, \textit{Asymptotic
structure of }$\mathcal{N}=2$ \textit{supergravity in\ 3D: extended super-BMS%
}$_{3}$ \textit{and nonlinear energy bounds}, JHEP \textbf{09} (2017) 030.
arXiv:1706.07542 [hep-th].

\bibitem{CCFR2} R. Caroca, P. Concha, O. Fierro, E. Rodríguez, \textit{%
Three-dimensional Poincaré supergravity and }$\mathcal{N}$\textit{-extended
supersymmetric BMS}$_{3}$ \textit{algebra}, arXiv:1812.05065 [hep-th].

\bibitem{CCRS} R. Caroca, P. Concha, E. Rodríguez, P. Salgado-Rebolledo,
\textit{Generalizing the }$\mathfrak{bms}_{3}$ \textit{and 2D-conformal
algebras by expanding the Virasoro algebra}, Eur. Phys. J. C \textbf{78}
(2018) 262. arXiv:1707.07209 [hep-th].

\bibitem{CRS} P.K. Concha, E.K. Rodríguez, P. Salgado, \textit{Generalized
supersymmetric cosmological term in N=1 Supergravity}, JHEP 08 (2015) 009.
arXiv:1504.01898 [hep-th].

\bibitem{CIRR} P.K. Concha, M.C. Ipinza, L. Ravera, E.K. Rodríguez, \textit{%
On the supersymmetric extension of Gauss-Bonnet like gravity}, JHEP \textbf{%
09} (2016) 007. arXiv:1607.00373 [hep-th].

\bibitem{BR} A. Baunadi, L. Ravera, \textit{Generalized AdS-Lorentz deformed
supergravity on a manifold with boundary}, arXiv:1803.08738 [hep-th].
\end{thebibliography}
\end{document}